\documentclass[11pt,twoside]{article}
\usepackage{asp2010}

\resetcounters

\begin{document}

\title{Detection of the 69$\, \mu$m band of crystalline forsterite in the Herschel MESS-program}
\author{B.L. de Vries$^{1}$, D. Klotz$^{2}$, R. Lombaert$^{1}$, A. Baier$^{2}$, J.A.D.L. Blommaert$^{1}$, L. Decin$^{1,3}$, F. Kerschbaum$^{2}$, W. Nowotny$^{2}$, T. Posch$^{2}$, H. Van Winckel$^{1}$, M.A.T. Groenewegen$^{4}$, T. Ueta$^{5}$, G. Van de Steene$^{4}$, B. Vandenbussche$^{1}$, P. Royer$^{1}$, C. Waelkens$^{1}$ \break
\affil{$^1$ Instituut voor Sterrenkunde, K.U. Leuven, Celestijnenlaan 200D, 3001 Leuven, Belgium}
\affil{$^2$University of Vienna, Department of Astronomy, T\"{u}rkenschanzstra\ss e 17, 1180 Vienna, Austria    }
\affil{$^3$Sterrenkundig Instituut Anton Pannekoek, University of Amsterdam, Science Park 904, 1098 XH, Amsterdam, The Netherlands    }
\affil{$^4$Koninklijke Sterrenwacht van Belgi\"{e}, Ringlaan 3, 1180 Brussels, Belgium}
\affil{$^5$Department of Physics and Astronomy, University of Denver, 2112 E. Wesley Ave., Denver, CO 80208, USA}
}

\begin{abstract}
In this article we present the detection of the 69$\, \mu$m band of the crystalline olivine forsterite within the MESS key program of Herschel. We determine the temperature of the forsterite grains by fitting the 69$\, \mu$m band.
\end{abstract}

\section*{Results}
Crystalline silicates are seen in many astronomical environments like disks around pre-main-sequence stars \citep{waelkens96, meeus01}, comets \citep{wooden02}, post-main-sequence stars \citep{waters96, mol02, devries10} and active galaxies \citep{kemper07}. The PACS instrument of Herschel allows us to study the 69$\, \mu$m band of forsterite with a better resolution and sensitivity than with ISO-LWS. Forsterite (Mg$_{2}$SiO$_{4}$) is the iron-poor end member of the solid solution of olivines. A broad range of evolved objects like AGB stars and planetary nebulae are observed in the MESS key program (Groenewegen et al submitted).

\begin{figure}
\begin{center}
 \includegraphics[scale = 0.5]{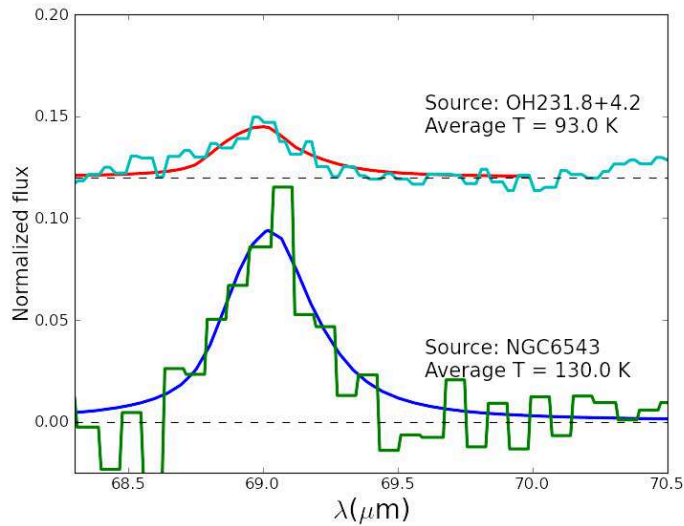} 
\end{center}
\caption{Fits to the 69$\, \mu$m band of forsterite using temperature dependent opacities \citep{suto06}. Fits are shown for the post-AGB star OH231.8+4.2 and the planetary nebula NGC 6543. From the fits we see that OH231.8+4.2 has somewhat colder forsterite than NGC 6543 }
\label{fits}
\end{figure}

Forsterite shows several spectral bands, for example, at 11.3$\, \mu$m and 33.6$\, \mu$m. But the band at 69$\, \mu$m is of particular interest because it is sensitive to the temperature of the crystalline olivine material. The central wavelength of the feature shifts to the red when the material has a higher temperature. Using temperature dependent opacities \citep{suto06,koike06} the 69$\, \mu$m band can be fitted in order to estimate the temperature of the circumstellar dust particles. Fig. \ref{fits} shows such a fit for the post-AGB star OH231.8+4.2 and the planetary nebula NGC 6543. These fits already show a different temperature for the dust grains in the envelopes of both objects.

The fits in Fig. \ref{fits} are made assuming crystalline olivine with no iron in its lattice structure. But when the lattice structure does contain some iron, the 69$\, \mu$m band also shifts to the red \citep{koike03, sturm10}. A shift due to a change in temperature of $\sim$100K can also be explained by including $\sim$1\% iron in the lattice structure of the crystalline olivine. So far there has not been any indication of crystalline olivines containing iron. By fitting the 69$\, \mu$m band in combination with mid-infrared features, we will be able to obtain compositional information of the crystalline olivines. 

\acknowledgements B.L. de Vries acknowledges support from the Fund for Scientific Research of Flanders (FWO) under grant number 6.0470.07. D. Klotz acknowledge funding by the Austrian Science Fund FWF under project number P19503 and P21988-N16, F. Kerschbaum under project numbers P18939-N16 and I163-N16 and W. Nowotny under project number P21988-N16.

\bibliography{devriesp}

\end{document}